\documentclass[a4paper,10pt,twoside]{cpc-hepnp}

\usepackage{multicol}
\usepackage{graphicx}
\usepackage{booktabs}
\usepackage{amssymb,bm,mathrsfs,bbm,amscd}
\usepackage[tbtags]{amsmath}
\usepackage{lastpage}

\begin{document}

\fancyhead[c]{Submitted to ``Chinese Physics C''}

\title{Velocity bunching in travelling wave accelerator with low acceleration gradient\thanks{Supported by National Natural Science Foundation of China (11205152) }}

\author{%
      HUANG Rui-Xuan
\quad HE Zhi-Gang$^{1)}$\email{hezhg@ustc.edu.cn}%
%\quad WANG Xiao-Jun(ÍõС¾ü)$^{1,2;2)}$\email{hepnp@mail.ihep.ac.cn}%
\quad LI Wei-Wei
\quad JIA Qi-Ka} \maketitle

\address{%
National Synchrotron Radiation Laboratory, University of Science and
Technology of China, Hefei, 230029, Anhui, China\\
%$^2$ {\bf Example}: Institute of High Energy Physics, Chinese Academy of Sciences, Beijing 100049, China\\
}

\begin{abstract}
 We present the analytical and simulated results concerning the influences of the acceleration gradient in the velocity bunching process, which is a bunch compression scheme that uses a traveling wave accelerating structure as a compressor. Our study shows that the bunch compression application with low acceleration gradient is more tolerant to phase jitter and more successful to obtain compressed electron beam with symmetrical longitudinal distribution and low energy spread. We also present a transverse emittance compensation scheme to compensate the emittance growth caused by the increasing of the space charge force in the compressing process that is easy to be adjusted for different compressing factors.
\end{abstract}



\begin{multicols}{2}

\section{Introduction}

In recent years, the demand for applications of high brightness - low emittance, high current, with sub-picosecond pulse length¡ªelectron beams has increased dramatically. In the fourth generation synchrotron light source community, high-brightness beams are needed for application to short wavelength free electron lasers (FEL), as well as for inverse-Compton-scattering (ICS) generation of short x-ray pulses. For studying novel accelerating techniques such as plasma-based accelerators and generation of coherent THz radiation, short electron bunches are also required.

Short bunches are commonly obtained by magnetic compression. In this scheme, the bunch is compressed when drifting through a series of dipoles arranged in a chicane configuration which can introduce an energy-dependent path length. Therefore an electron bunch with the proper time-energy correlation can be shortened in the chicane. The time-energy correlation along the bunch can be tuned by means of an accelerating section upstream from the chicane. Great progress has been made in this this field, but magnetic compression may introduce momentum spread and transverse emittance dilution due to the bunch self-interaction via coherent synchrotron radiation\cite{lab1}.  To obtain a smaller and more symmetrical electron beam, a linear energy-time correlation is required along the bunch, which can be realized by a accelerating structure at a higher harmonic\cite{lab2}, with respect to the main accelerating linac RF.

Velocity bunching relies on the phase slippage between the electrons and the rf wave that occurs during the acceleration of nonultrarelativistic electrons. It was experimentally observed in photocathode rf gun\cite{lab3} and proposed to integrate the velocity bunching scheme in the next photoinjector designs using a dedicated rf structure downstream of the rf electron source\cite{lab4}. Previous experimental works showed the compression ability of the velocity bunching method\cite{lab5,lab6}. Furthermore, the emittance growth in the compressing process was completely compensated by long solenoids\cite{lab7} when the compression factor is 3.

This paper mainly focuses on the beam bunching in the traveling wave structure with low acceleration gradient (4~MV/m) instead of high acceleration gradient (normally 20~MV/m) in previous work. A brief analysis of the velocity bunching mechanism is presented in the Section 2 firstly. In Section 3, the analytical and simulated results of bunch compression within high and low acceleration gradient accelerators are described, and the conclusion is figured out that  a traveling wave accelerating structure with low acceleration gradient is more tolerant to phase jitter and easier to obtain compressed electron beam with symmetrical longitudinal distribution and low energy spread in the velocity bunching process. A transverse emittance compensation scheme is shown in Section 4, which is easy to be adjusted for different compressing factors. Section 5 presents a summary of this paper.

\section{Velocity bunching mechanism}

In the velocity bunching process, the longitudinal phase space rotation is based on a correlated time-velocity chirp in the electron bunch, so that electrons on the tail of the bunch are faster than electrons in the bunch head. This rotation occurs inside a traveling rf wave of a long multicell rf structure which applies an off crest energy chirp to the injected beam as well as accelerates it.. This is possible if the injected beam is slightly slower than the phase velocity of the rf wave so that when injected at the zero crossing field phase it slips back to phases where the field is accelerating, but is simultaneously chirped and compressed.

An electron in an rf traveling wave accelerating structure experiences the longitudinal electric field:
\begin{eqnarray}
{E_z} = {E_0}\sin (\phi )
\end{eqnarray}
where ~$E_0$~is the peak field, ~$\phi  = kz - \omega t + {\phi _0}$~is the phase of the electron with respect to the wave and ~$\phi _0$~is the injection phase of the electron with respect to the rf wave. The evolution of~$\phi$~can be expressed as a function of z:
\begin{eqnarray}
\frac{{d\phi }}{{dz}} = k-\omega \frac{{dt}}{{dz}} = k-\frac{\omega }{{\beta c}} = k(1-\frac{\gamma }{{\sqrt {{\gamma ^2} - 1} }})
\end{eqnarray}
The energy gradient can be written as\cite{lab8}:
\begin{eqnarray}
\frac{{d\gamma }}{{dz}} = \alpha k\sin (\phi )
\end{eqnarray}
where~$\alpha  \equiv e{E_0}/m{c^2}k$~is defined as dimensionless vector potential amplitude of the wave. The equations (2) and (3) with the initial conditions~${\gamma _{z = 0}} = {\gamma _0}$~and~${\phi _{z = 0}} = {\phi _0}$~describe the longitudinal motion of an electron in the rf structure. Using a separation of variables approach, one can get
\begin{eqnarray}
\alpha \cos \phi  + \gamma  - \sqrt {{\gamma ^2} - 1}  = C
\end{eqnarray}
the~$\phi$~can be expressed as a function of~$\gamma$~:
\begin{eqnarray}
\phi (\gamma ) = \arccos (\frac{{C - \gamma  + \sqrt {{\gamma ^2} - 1} }}{\alpha })
\end{eqnarray}
where the constant C is set by the initial conditions of the problem:~$C=\alpha \cos \phi_0  + \gamma_0  - \sqrt {{\gamma_0 ^2} - 1}$~.
The final phase of electron at the exit of the accelerator is
\begin{eqnarray}
\phi_e (\gamma_e ) = \arccos (\frac{{C - \gamma_e  + \sqrt {{\gamma_e ^2} - 1} }}{\alpha })
\end{eqnarray}
If the~$\gamma_e$~is high enough,~$\sqrt {{\gamma_e ^2} - 1}- \gamma_e\simeq0$~. With the approximation~$\gamma_0-\sqrt {{\gamma_0 ^2} - 1}\simeq 1/(2\gamma_0)$~. Then, the final phase becomes:
\begin{eqnarray}
\phi_e = \arccos (\cos\phi_0+1/(2\alpha\gamma_0))
\end{eqnarray}
Expanding Eq. (7) to first order in the initial energy spread and initial phase spread gives
\begin{eqnarray}
\Delta {\phi _e} = \frac{{\sin {\phi _0}}}{{\sin {\phi _e}}}\Delta {\phi _0} + \frac{1}{{2\alpha \gamma _0^2\sin {\phi _e}}}\Delta {\gamma _0}
\end{eqnarray}
Hence depending upon the incoming energy and phase extents (initial bunch length), the phase of injection in the rf structure~$\phi_0$~can be tuned to minimize the phase extent of extraction (final bunch length).
\section{Bunch compression in low and high gradient accelerators}
Based on equations (6) and (8), one can find that the compression factor is less sensitive to the injection phase~$\phi_0$~and more tolerant to the phase jitter when the peak field~$E_0$~is lower. Fig. 1 shows the simulated phase scanning results at different acceleration gradients by using the code ASTRA\cite{lab9}.
\begin{center}
\includegraphics[width=8.0cm]{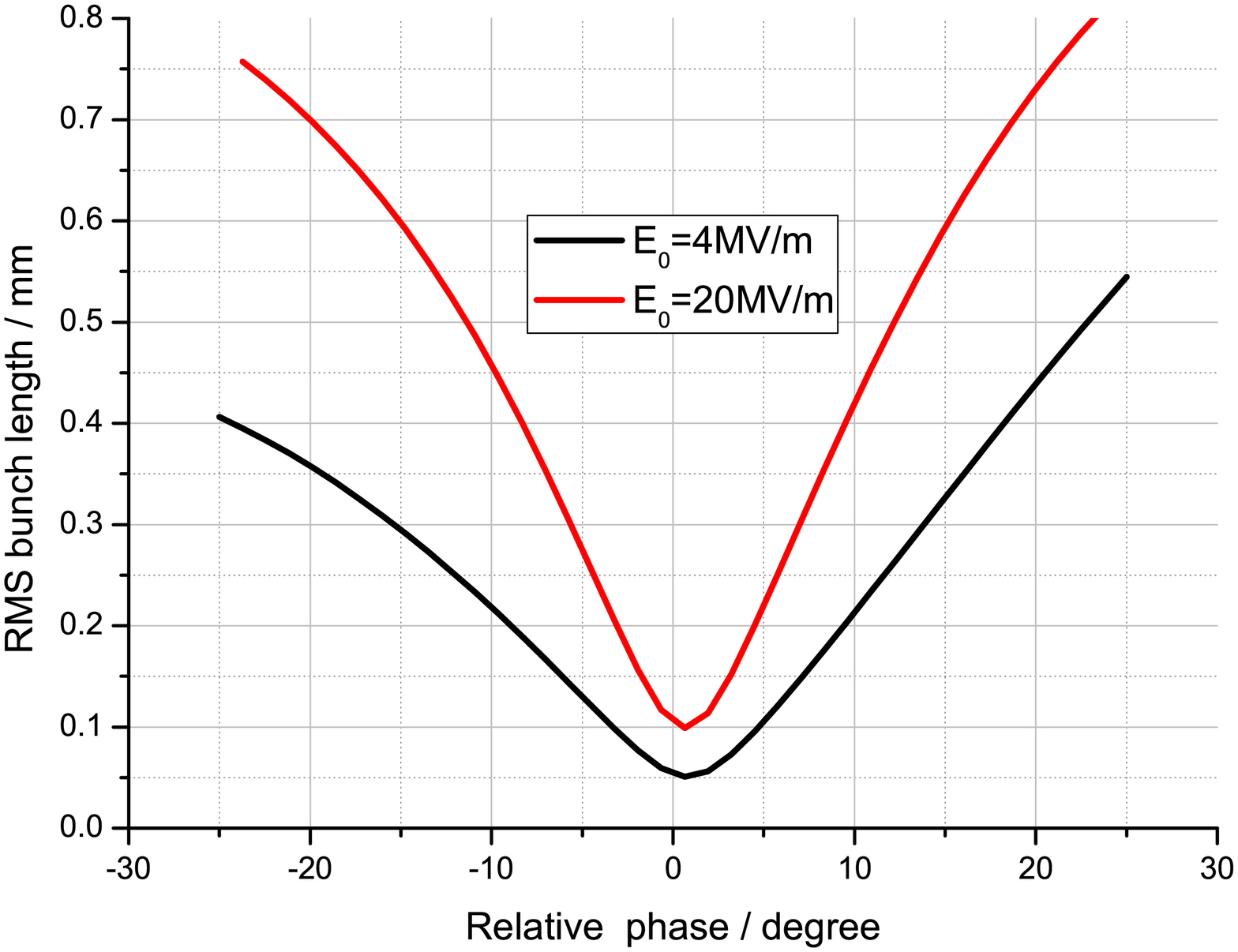}
\figcaption{\label{fig1} The rms bunch length as a function of relative phase (There is about 28.7 degrees phase shift between the two results. In the figure, we do a translation for contrast).}
\end{center}
The simulation shows a coincident result to the analytical conclusion. It indicates that higher precision phase jitter control and power supply for the solenoids to compensate the transverse emittance growth (The emittance compensation scheme will be presented in next section) are needed in experimental practice when the acceleration gradient is high. A lower acceleration gradient in the velocity bunching process has been confirmed to be more tolerant to phase jitter and power precision. Furthermore, with a lower gradient, the compressed electron beam tends to develop more symmetrical longitudinal distribution. Analytical and simulated results will be presented in following paragraph.

The velocity bunching process can be shown as Fig. 2. To compress the electron bunch, the injection phase should be set at the phase that is tens degrees off the crest. The velocity of the initial beam is smaller than the rf wave (~$\beta_\phi=1$~), there is phase slippage between the initial and compressed beam. For a high gradient accelerator, the phase slippage is smaller, but to obtain the same compression factor as in the low gradient accelerator, the injection phase should be closer to the crest; when the injection phase is farther from the crest, one can get a compressed beam with a single spike and most of electrons are in the spike, but the compression factor is too large, the emittance compensation is difficult at present. In this paper, the accelerators are the normal S-band (~$2856~MHz$~),~$2/3\pi$~mode,~$3~m$~long travelling wave accelerators. One also can set an appropriate injection phase to make the phase slippage in the range of linearity within a short accelerator, then the electron bunch is imposed a more linear negative energy chirp (the speed of electrons in the head is slower than the speed of electrons in the tail), and the electron bunch can be compressed enormously in a drift space with appropriate length, which has been presented in Reference\cite{lab6} and named by the authors as ``ballistic bunching''.
\begin{center}
\includegraphics[width=8.0cm]{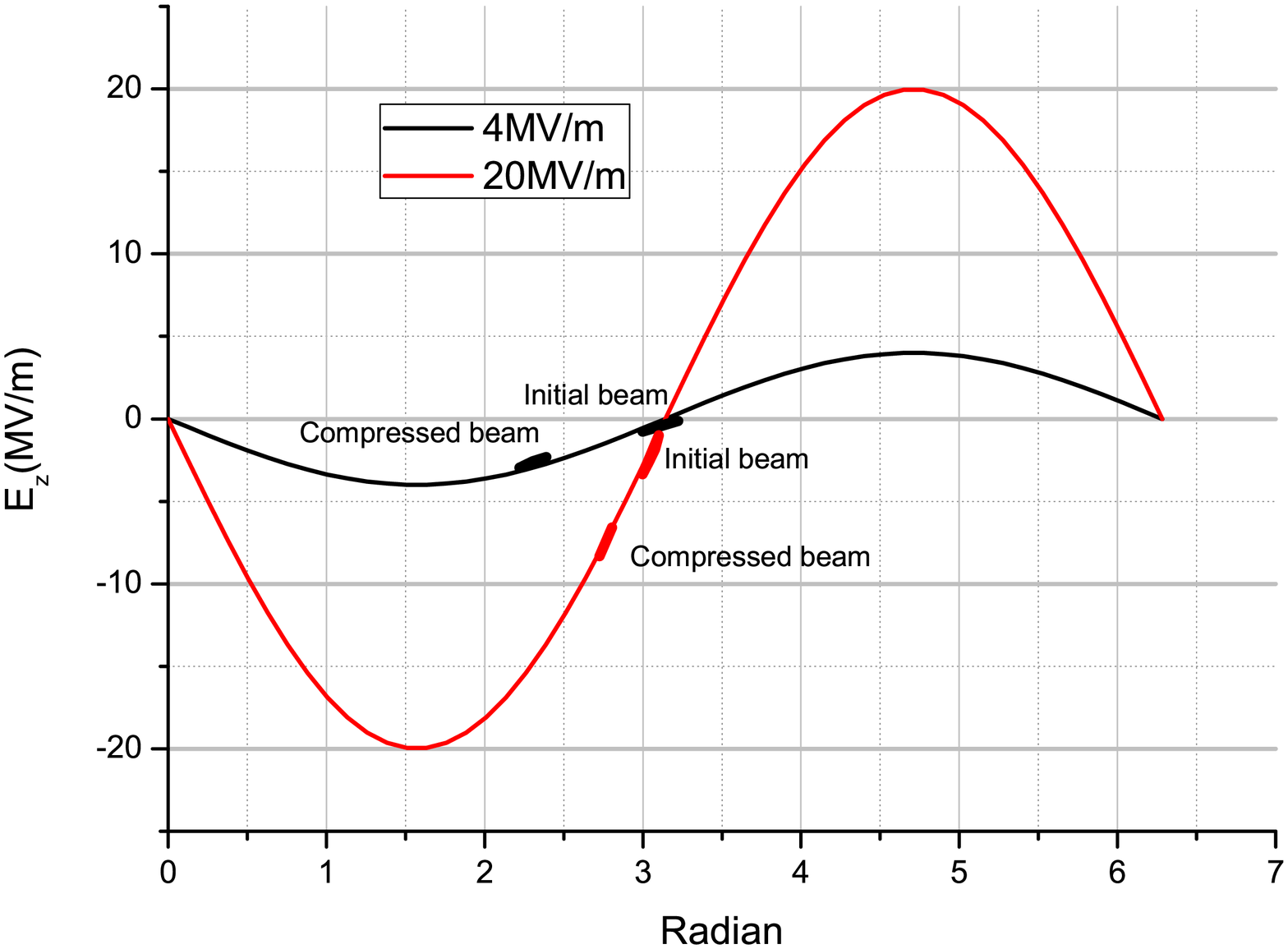}
\figcaption{\label{fig2} Velocity bunching process.}
\end{center}
Eq. (1) can be expanded as:
\begin{eqnarray}
{E_z} = {E_0}(\phi  - \frac{{{\phi ^3}}}{{3!}} + \frac{{{\phi ^5}}}{{5!}} - \frac{{{\phi ^7}}}{{7!}} + \frac{{{\phi ^9}}}{{9!}} \cdots )
\end{eqnarray}
If there is no higher order terms in the equation, the longitudinal distribution state of the initial beam  can be preserved during the velocity bunching process when the initial energy spread is very small or it can be treat as a linear one. Unfortunately, there are always higher order terms in the equation. So£¬reducing the~$E_0$~may be an available way to prevent the longitudinal distortion. By using the equations (2) and (3), the numerical calculation results show the rationality of the analysis. However, the space charge force and magnetic force are not considered. In the following content, the simulation results will be presented.

Longitudinal distribution after compression by different~$E_0$~should be compared. The current profiles of the initial bunch and the compressed bunches (with almost identical bunch length) are shown in Fig. 3.
\begin{center}
\includegraphics[width=8.0cm]{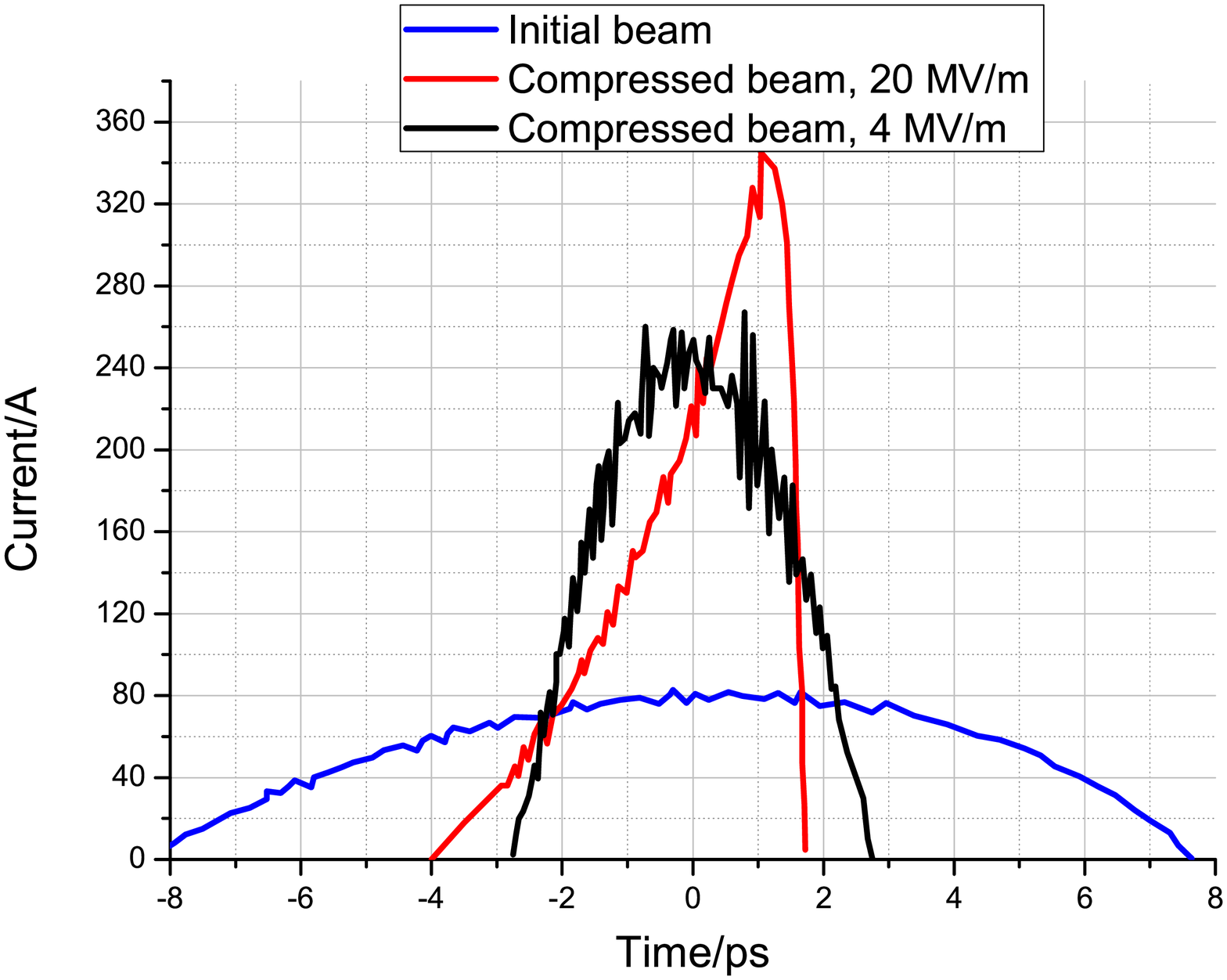}
\figcaption{\label{fig3}The current profiles of the initial bunch (blue line) and the compressed bunches (black line: compressed bunch in~$4~MV/m$~gradient accelerator; red line: compressed bunch in~$20~MV/m$~gradient accelerator).}
\end{center}
From the above figure£¬it appears that a traveling wave accelerating structure with low acceleration gradient is more successful to obtain compressed electron with symmetrical longitudinal distribution.
Fig. 4 shows the bunch length evolutions along the beam direction when the acceleration gradients are~$E_0=4~MV/m$~and~$E_0=20~MV/m$~.
\begin{center}
\includegraphics[width=8.0cm]{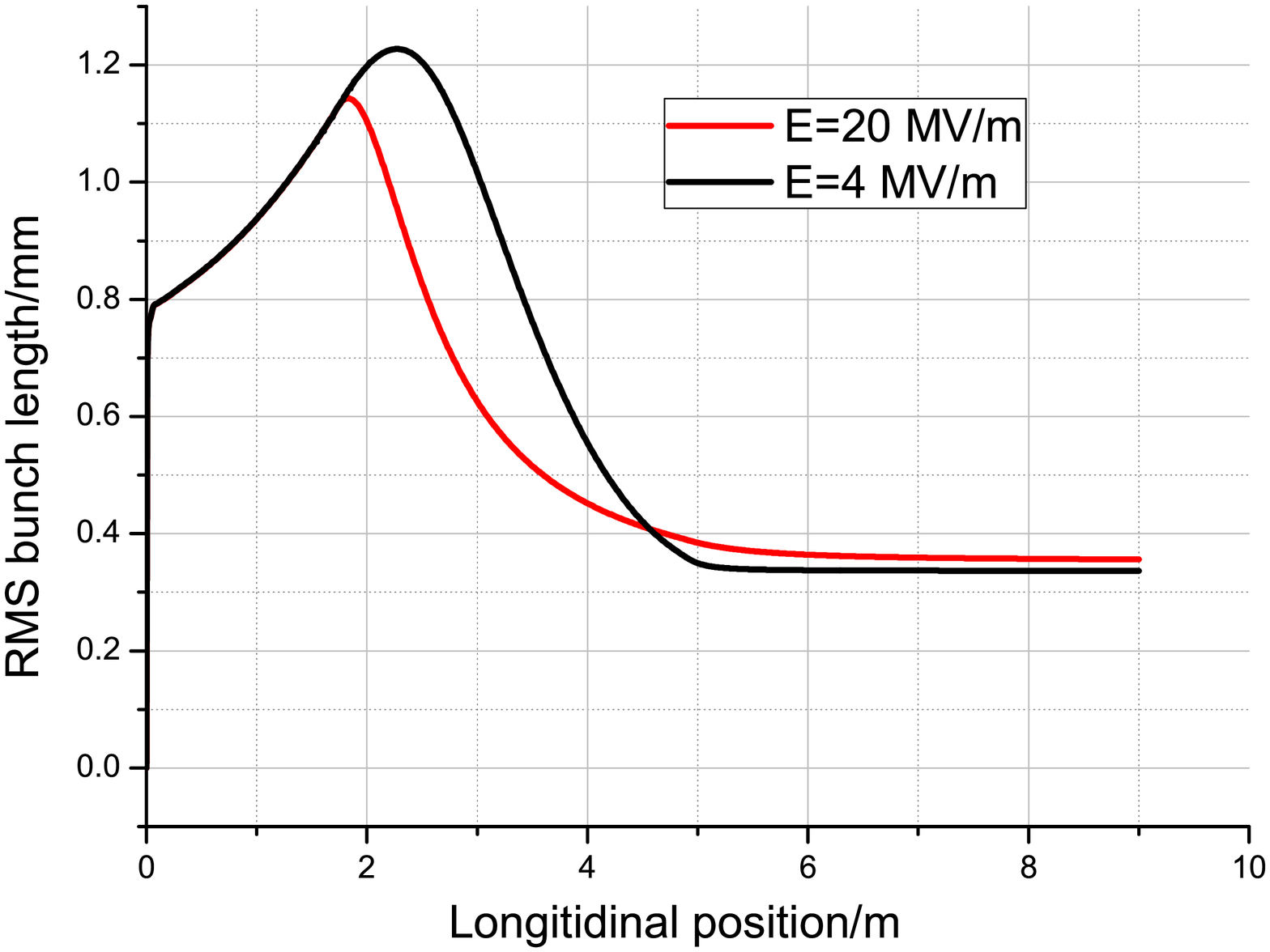}
\figcaption{\label{fig4}RMS bunch length evolution along the longitudinal position.}
\end{center}

As we know that to avoid the longitudinal distortion in magnetic compression, an X-band accelerator system is normally used to linearize the energy chirp of the electron beam\cite{lab2}, which is cost almost one million dollar. About this problem in the velocity bunching process, theoretical and experimental works are worth doing in detail.

When the~$E_0$~is higher, the energy spread in the bunch is greater, because the bunch compression is achieved by the velocity difference within the bunch. Fig. 5 shows a contrast of the energy spread evolution when the gradient are ~$E_0=4~MV/m$~and~$E_0=20~MV/m$~.
\begin{center}
\includegraphics[width=8.0cm]{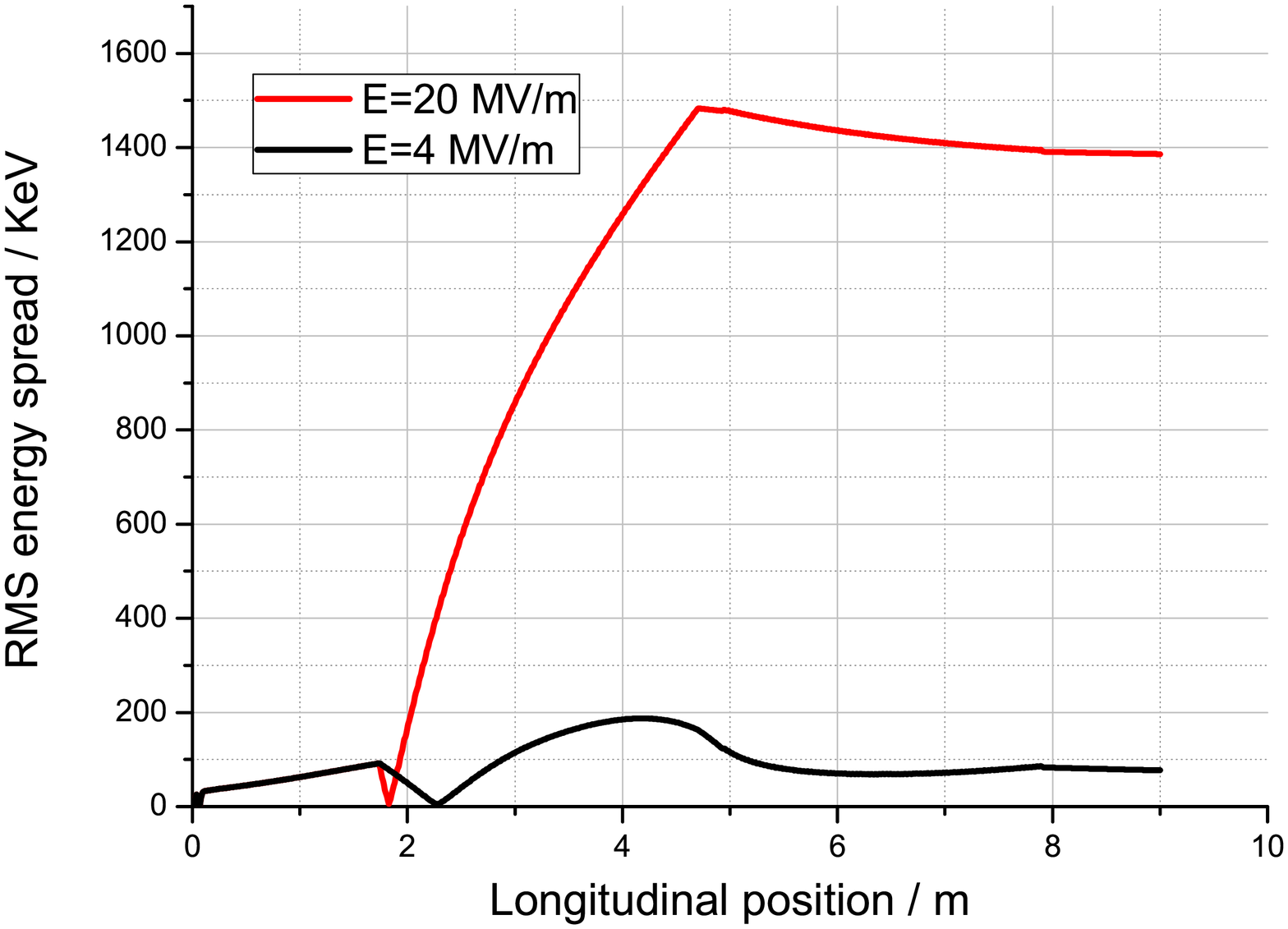}
\figcaption{\label{fig5}RMS energy spread evolution at different acceleration gradients.}
\end{center}

\section{Emittance compensation}
In this section, an emittance compensation scheme will be suggested during the velocity bunching process.
\begin{center}
\includegraphics[width=8.0cm]{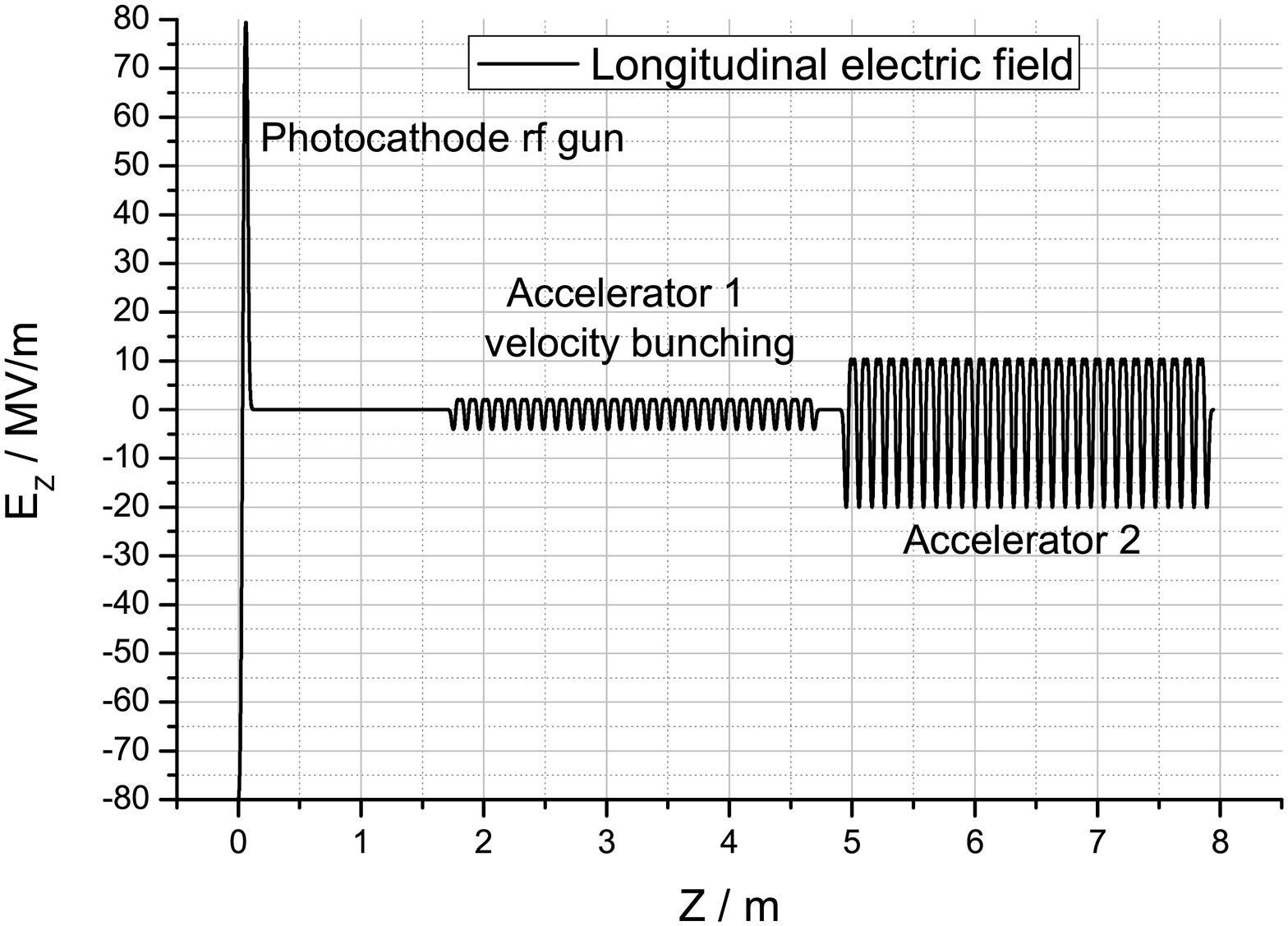}
\figcaption{\label{fig6}The electric field of accelerators along the beam line.}
\end{center}

The electric and magnetic fields along the beam line are shown in Fig. 6 and Fig. 7. The peak acceleration gradients of the photocathode rf gun, accelerator 1 and accelerator 2 are~$80~MV/m$~,~$4~MV/m$~and~$20~MV/m$~respectively. Four solenoids are used for emittance compensation, ~$B_0$~is located at~$10~cm$~downstream from the photocathode, and the other three are around the accelerators.
\begin{center}
\includegraphics[width=8.0cm]{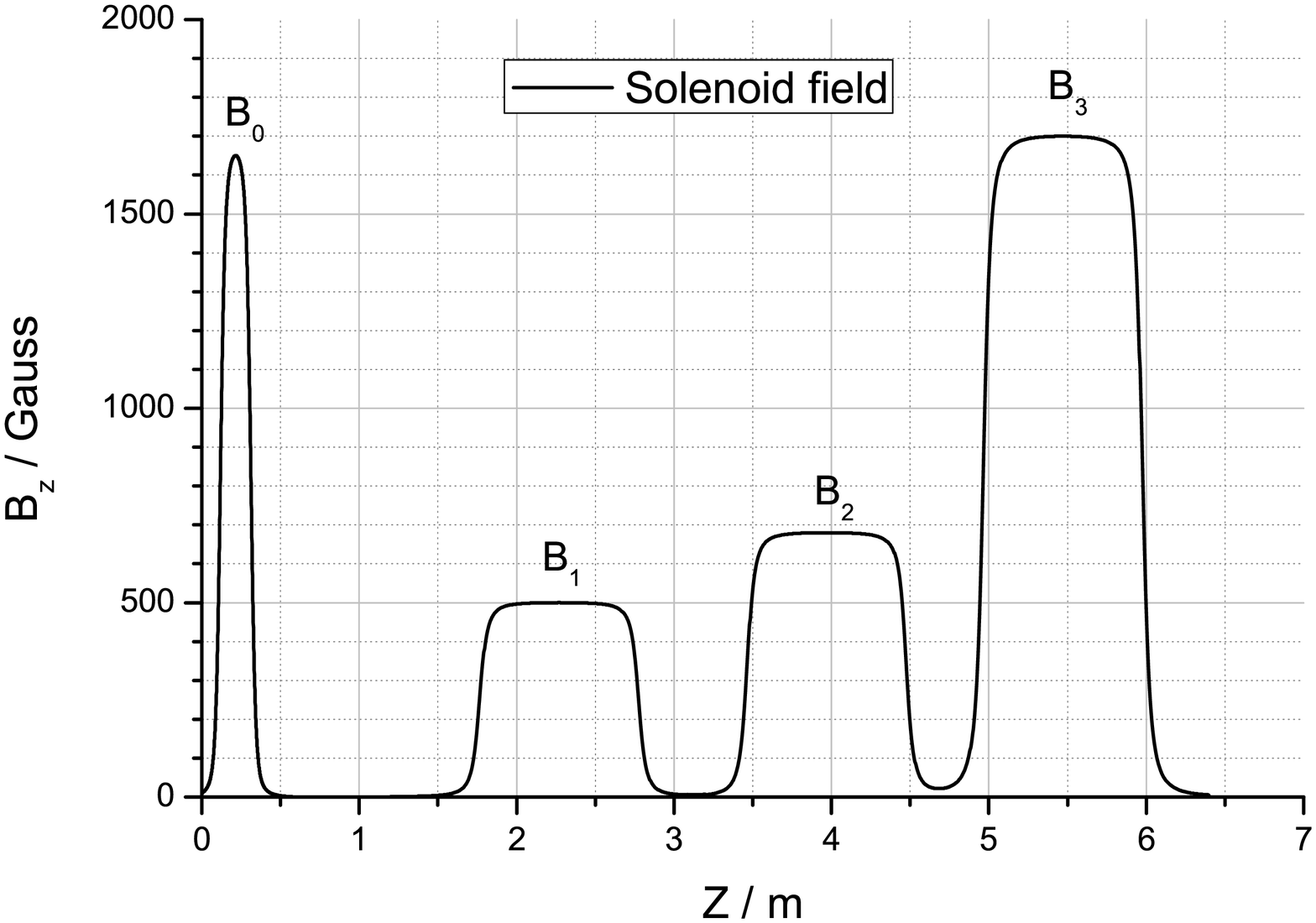}
\figcaption{\label{fig7}The magnetic field of solenoids along the beam line.}
\end{center}

To prevent irreversible emittance growth during bunch compression, the crux is to preserve the laminarity of the electron beam with an envelope propagated as close as possible to a Brillouin-like flow, represented by an invariant envelope\cite{lab10} as generalized to the context of beam compression and thus increasing~$I$~during acceleration. Mismatches between the space charge forces and the external focusing gradient produce slice envelope oscillations that cause normalized emittance oscillations. It has been shown that in order to keep such oscillations under control during the velocity bunching, the beam has to be injected into the accelerator with a laminar envelope waist (~$\sigma ^{'} = 0$~) and the envelope has to be
matched to the accelerating and focusing gradients so that it can stay close to an equilibrium mode\cite{lab10}. Long solenoids around the accelerator are used to provide the required focusing.

The transverse emittance and beam size evolutions along the beam direction are shown in Fig. 8. For electron bunches with different compression factors at different injection phases, one just need to scan the strength of the solenoids to compensate the emittance growth as shown in Fig. 8.
\begin{center}
\includegraphics[width=8.0cm]{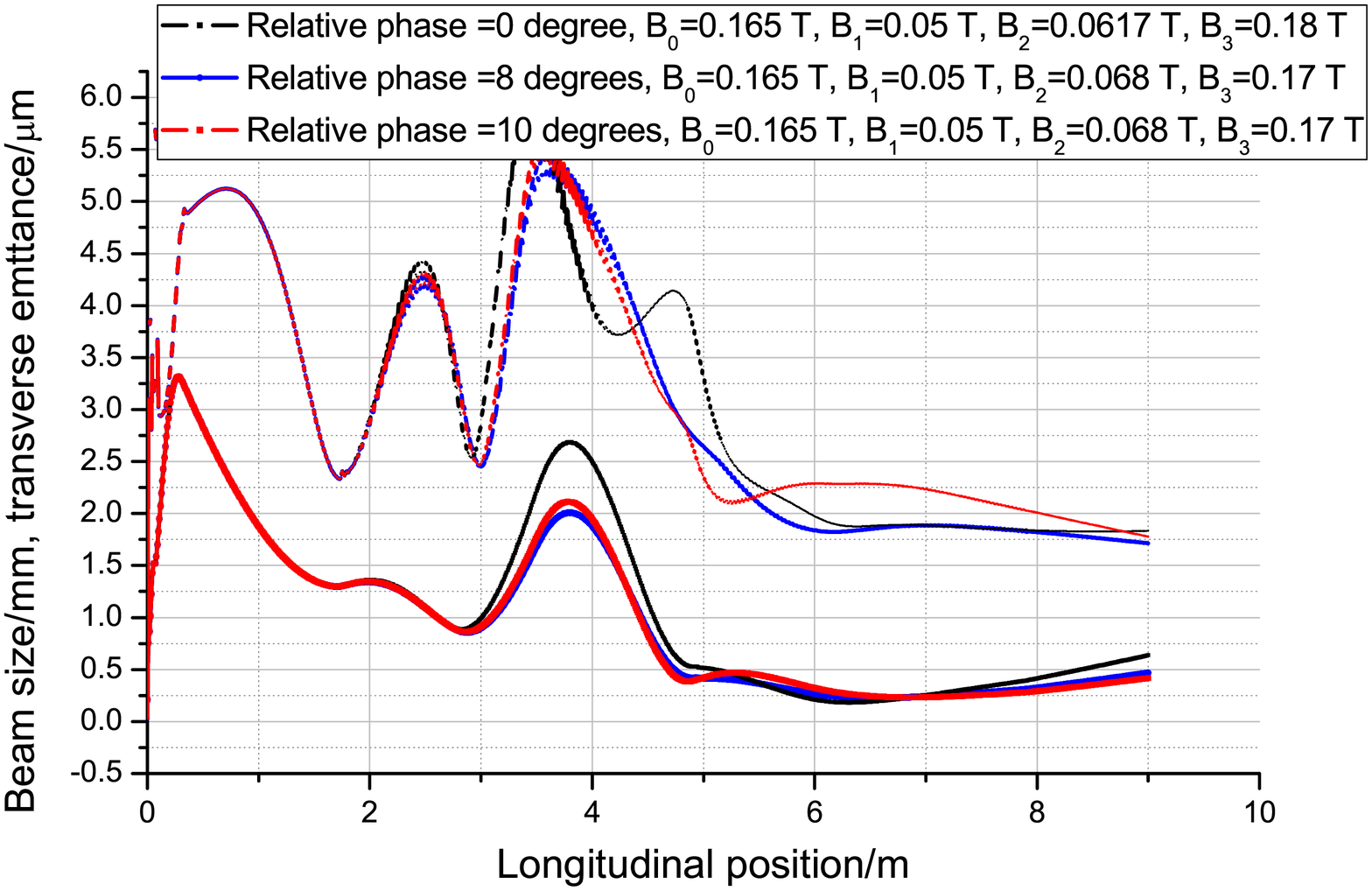}
\figcaption{\label{fig8}The transverse emittance and beam size evolutions along the beam direction at different injection phases, the optimal magnetic strengths of solenoids are also shown in the figure.}
\end{center}
 Fig. 9 shows the current profiles for different compressed beam at different injection phases. We should point out that the magnetic force of long solenoids (around the accelerator) can affect the compression, which can be found by contrasting the results in Fig. 9 and phase scanning result at~$E_0 = 4~MV/m$~without the long solenoids in Fig. 1.
 \begin{center}
\includegraphics[width=8.0cm]{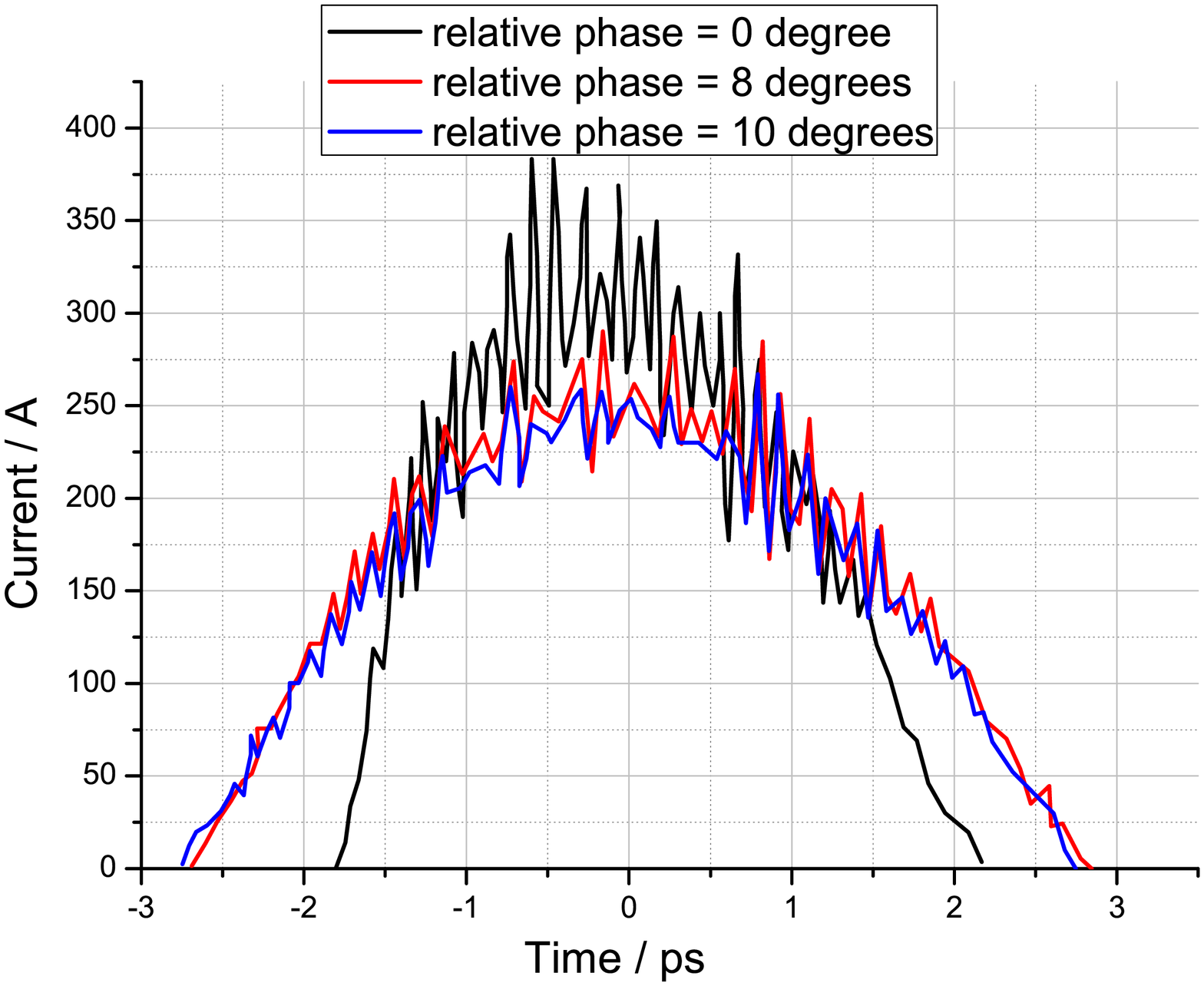}
\figcaption{\label{fig9}The current profiles for different compressed beam at different injection phases.}
\end{center}
\section{Summary and conclusion}
Based on the analyses, we contrast the ASTRA simulated results of velocity bunching process in travelling wave structure with low (~$4~MV/m$~) and high (~$20~MV/m$~) acceleration gradient. According to the results, the conclusion is figured out that the bunch compression application with low acceleration gradient is more tolerant to phase jitter and should be useful for obtaining better performance beams with symmetrical longitudinal distribution and low energy spread. Furthermore£¬a successful improvement of transverse emittance during compression is possible with optimized long solenoids£¬which is easily satisfied for different compressing factor.

\acknowledgments{The authors thank Prof. PEI, Yuan-Ji for insightful discussions.}
\end{multicols}

%\vspace{10mm}

%\begin{multicols}{2}

%\subsection*{Appendices A}
%\begin{small}

%\noindent{\bf Subtitle}

%\end{small}
%\end{multicols}

\vspace{-1mm}
\centerline{\rule{80mm}{0.1pt}}
\vspace{2mm}

\begin{multicols}{2}

\end{multicols}

\clearpage

\end{document}